\documentclass[twocolumn,prl,showpacs,preprintnumbers,amsmath,amssymb]{revtex4}
\usepackage{graphicx}% Include figure files
\usepackage{dcolumn}% Align table columns on decimal point
\usepackage{bm}% bold math
\usepackage{subfigure}

\begin{document}

\title{Stabilization of Skyrmion textures by
 uniaxial distortions in noncentrosymmetric cubic helimagnets
}

%\author{A. B. Butenko$^{1,2}$, A. A. Leonov$^{1,2}$ ,  U. K. R\"o\ss ler$^{1}$, A. N. Bogdanov$^{1}$}
\author{A. B. Butenko$^{1,2}$}
\author{A. A. Leonov$^{1,2}$}\thanks{Corresponding author:  +49-351-4659-385; fax:  +49-351-4659-490; E-mail address: a.leonov@ifw-dresden.de}%\footnotemark[1] %\thanks{Corresponding author }
\author{U. K. R\"o\ss ler$^{1}$} \thanks{Tel:+49-351-4659-542; E-mail address: u.roessler@ifw-dresden.de }%\footnotemark[2]

\author{A. N. Bogdanov$^{1}$}

%\footnotetext[1]{
%Tel: +49-351-4659-385; fax:  +49-351-4659-490;  
%E-mail address: a.leonov@ifw-dresden.de}

%\footnotetext[2]{
%Tel: +49-351-4659-542; fax:  +49-351-4659-490;  
%E-mail address: u.roessler@ifw-dresden.de}

\address{$^1$IFW Dresden, Postfach 270116, D-01171 Dresden, Germany}

\address{$^2$Donetsk Institute for Physics and Technology, 
R. Luxemburg 72, 83114 Donetsk, Ukraine}
\date{\today}

\begin{abstract}
{
In cubic noncentrosymmetric
ferromagnets uniaxial distortions  
suppress the helical states  and stabilize 
Skyrmion lattices in a broad range
of thermodynamical parameters.
Using a phenomenological theory for modulated and
localized states in chiral magnets, the equilibrium parameters of
the Skyrmion and helical states are derived as functions of the
applied magnetic field and induced uniaxial anisotropy.

These results show that due to a combined effect of 
induced  uniaxial anisotropy and an applied magnetic field
Skyrmion lattices can be formed as thermodynamically stable states 
in large intervals of magnetic
field and temperatures in cubic helimagnets,
e.g., in intermetallic compounds MnSi, FeGe, 
(Fe,Co)Si.
We argue that this mechanism is responsible for the formation
of Skyrmion states recently observed in thin layers of 
Fe$_{0.5}$Co$_{0.5}$Si [X.Z.Yu et al., Nature \textbf{465}(2010) 901].
}
\end{abstract}

\pacs{
75.30.Kz 
% Magnetic phase boundaries (including magnetic transitions, metamagnetism, etc.)
75.10.-b 
% General theory and models of magnetic ordering 
% (see also 05.50 Lattice theory and statistics)
75.70.-i,
%Magnetic properties of thin films, surfaces, and interfaces 
% (for magnetic properties of nanostructures, see 75.75.+a)
}
% %%% PACS numbers

%\keywords{ }%Use showkeys class option if keyword
                              %display desired
         
\maketitle

% \clearpage
%

\vspace{5mm}

Multidimensional localized and modulated 
structures (Skyrmions) are  intensively
investigated in many areas of physics.
\cite{Schmeller95,Nature06}
In the majority of nonlinear field models,
Skyrmionic states appear only as dynamic
excitations, but static configurations are
generally unstable and collapse spontaneously
into topological singularities.\cite{Derrick}
These instabilities can be overcome 
if the energy functionals contain
(i) contributions with higher-order 
spatial derivatives (\textit{Skyrme mechanism}),\cite{Skyrme61}
or (ii)  terms linear with respect to spatial
derivatives (so called Lifshitz invariants) 
\cite{JETP89,PRL01}
\begin{eqnarray}
\Lambda_{ij}^{(k)} = L_i\partial_k L_j-L_j\partial_k L_i,
\label{Lifshitz}
\end{eqnarray}
where $\mathbf{L}$ is a vector
order parameter (e.g. the magnetization
vector $\mathbf{M}$ in magnetic materials or
the director $\mathbf{n}$ in chiral liquid crystals), 
$\partial_k L_i \equiv \partial L_i/\partial x_k$
are spatial derivatives of the order parameter.

\begin{figure*}
\includegraphics[width=17cm]{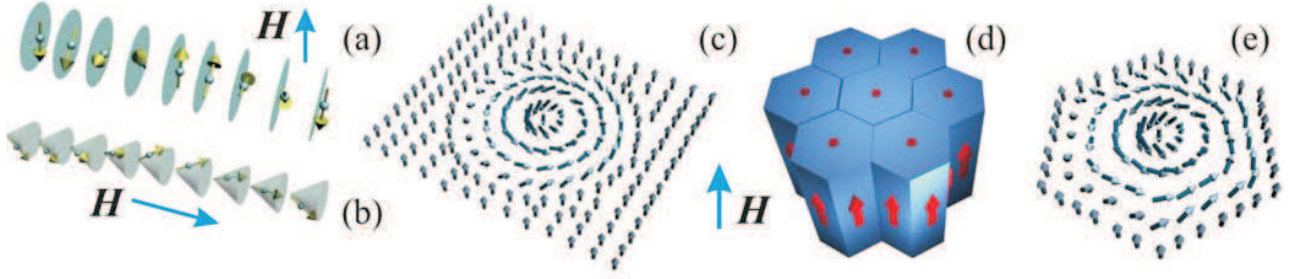}
\caption{ 
(Color online) Chiral modulated states in noncentrosymmetric cubic
magnets: the \textit{helicoid}, a distorted helix with the propagation
direction perpendicular to the applied field (a),
the \textit{conical} helix propagates along the applied field (b),
an isolated Skyrmion (c), a hexagonal Skyrmion lattice (d)
with the internal structure of the unit cell (e).
\label{Fig1}
}
\end{figure*}

In condensed matter physics
there are no physical interactions
underlying  energy contributions 
with  higher-order spatial derivatives.\cite{Remark1}
On the contrary, the invariants of
type (\ref{Lifshitz}) arise in systems
with intrinsic \cite{Dz64,Degennes93} and
induced chirality.\cite{PRL01}
Particularly, in noncentrosymmetric
magnetic materials such interactions
stem from the chiral part of spin-orbit
couplings (Dzyaloshinskii-Moriya interactions).\cite{Dz64}
Chiral interactions of type (\ref{Lifshitz})
stabilize helical \cite{Dz64,Bak80}
and Skyrmionic structures \cite{JETP89,JMMM94}
with fixed rotation sense (Fig. \ref{Fig1}).
Theoretically, isolated Skyrmions
and Skyrmion lattices have been investigated
for several classes of noncentrosymmetric
systems (e.g. see Refs.\  \onlinecite{Nature06,PRL01,JMMM94}
and bibliography in Ref.\  \onlinecite{Nature06}).
%
%Recent observations of specific precursor effects
%in cubic helimagnet MnSi \cite{Pappas09} together
%with earlier experimental findings 
%\cite{Lebech95}
%support the theoretical predictions of Skyrmion
%states from Refs.~$[$\onlinecite{Nature06,JMMM94}$]$. 
%
Contrary to uniaxial chiral ferromagnets
from Laue classes C$_{nv}$ and D$_{2d}$
where thermodynamically stable 
Skyrmionic states exist in a broad range
of applied magnetic fields and temperatures,
\cite{JMMM94,Nature06,condmat2010}
Skyrmionic states compete with one-dimensionally
modulated helices in cubic helimagnets.\cite{Nature06}
Hence, in the main part of the magnetic phase diagram 
for cubic helimagnets,
the ordered state helical structures 
correspond to the global energy minima 
(such texture have been recently observed in (Fe,Co)Si
alloys  \cite{Uchida06} and in FeGe \cite{Uchida08}.
Therefore, additional effects are necessary
to stabilize Skyrmionic states 
in these systems.\cite{Nature06,condmat2010}
In this paper we demonstrate 
that uniaxial distortions suppress 
the helical phases and enable the thermodynamic
stability of the Skyrmion lattice in a broad range
of applied magnetic fields. 
The calculated magnetic
phase diagram allows to formulate practical recommendations
on the possibility to stabilize Skyrmion states 
at low temperatures in MnSi, FeGe, (Fe,Co)Si 
and similar intermetallic compounds with B20-structure.

Following the phenomenological theory developed
in Refs.~\cite{Dz64,Bak80} we write 
the magnetic energy density for a cubic helimagnet
with uniaxial distortions along $z$-axis as
\begin{equation}
w=A \left(\mathbf{grad} \mathbf{M}\right)^2
-\mathbf{M}\cdot\mathbf{H}+w_D+ w_a -K M_z^2\,,
\label{density}
\end{equation}
where $A$ is the exchange stiffness, 
the second term is the Zeeman energy,
$w_D=D\,(\Lambda_{yx}^{(z)}+\Lambda_{xz}^{(y)}+\Lambda_{zy}^{(x)})$
=$D\,\mathbf{M}\cdot \mathrm{rot}\mathbf{M}$ is the
chiral energy with the Dzyaloshinskii constant $D$,
$w_a = \sum_i[ B(\partial_i M_i)^2 + K_c M_i^4]$ includes
exchange ($B$) and cubic ($K_c$) anistropies
\cite{Bak80}. The last term in (\ref{density}) is uniaxial
anisotropy induced by distortions.

The Dzyaloshinskii-Moriya energy $w_D$ (\ref{density}) 
favours spatially modulated chiral states
where the magnetization rotates with a fixed turning sense 
in the plane perpendicular to 
the propagation direction (Fig.~\ref{Fig1}).
The sign and magnitude of the Dzyaloshinskii constant $D$ 
determine the modulation
period and the sense of rotation, respectively.
Thus, in zero magnetic field, $\mathbf{H}$ = 0, and 
for zero anisotropies, $B = K_c = K$ = 0, a flat helix forms the 
magnetic ground state
as a \textit{single harmonic} mode with wave number $q_0 = D/(2A)$, 
where the phase angle $\phi$ of the magnetization varies linearly
along the propagation direction $\xi$, $\phi (\xi) = \xi q_0$.\cite{Dz64,Bak80}
Intrinsic cubic anisotropy $w_a$ is
much weaker than the energy terms in Eq.~(\ref{density})
and are neglected in further calculations. Its role will
be discussed to the end of the paper.
The solutions for one-dimensional modulations include (i) distorted
helices (\textit{helicoids}) 
(Fig. \ref{Fig1} (a)) and (ii) 
\textit{conical} phases, helices with the propagation
vector along the applied field (Fig. \ref{Fig1} (b)).

For the latter state the equilibrium parameters are
readily derived in analytical form
\begin{eqnarray}
 \cos \theta = \frac{H}{H_0},\; \psi = \frac{z}{L_D},\;
 H_0 =H_d \left(1- \frac{K}{K_0}\right), 
\label{conical}
\end{eqnarray}
and the equilibrium energy density 
$W_C = -K_0 M^2[H^2/(H_0 H_d)- 1]$.
In Eqs. (\ref{conical})
$L_D = 2 A/D$ is the characteristic
length unit of the modulated states.
In the critical field $H_0 (K)$ the conical 
helix flips into the saturated state.
The characteristic field $H_d=D^2\,M\,/\,(2A)$ is
the flip field for zero distortions. 
The anisotropy value $K_0=H_d/(2M)$ marks the 
critical value for uniaxial distortions 
suppressing the conical phase in zero field.
The analytical solutions for helicoids
have been derived by Dzyaloshinskii.\cite{Dz64}

In addition to the helical phase the model
Eq.~(\ref{density}) has solutions for two-dimensional
modulated states (Skyrmions).\cite{JETP89,JMMM94}
To describe Skyrmionic states in
a magnetic field along the $z$-axis
we introduce spherical coordinates for
the magnetization vector
$\mathbf{M}=M(\sin\theta\cos\psi;
\sin\theta\sin\psi;\cos\theta))$
and the cylindrical coordinates for
the spatial variable
$\textit{\textbf{r}}=L_D 
(\rho\cos\varphi;\rho\sin\varphi;z)$.
Minimization of energy (\ref{density})
yields rotationally symmetric solutions 
$\varphi=\psi+\pi/2$  and
$\theta=\theta(\rho)$ derived from equation
\begin{align}
&\frac{1}{\rho} \left(\rho \theta_{\rho \rho} 
+\theta_{\rho} -\frac{\sin \theta \cos \theta}{\rho} \right) 
+\frac{2}{\rho} \sin^2 \theta -f(\theta) = 0,\nonumber\\
&f(\theta)=(K/K_0)\sin\theta \cos\theta+ (H/H_d) \sin\theta,
\label{DifEq}
\end{align}
with boundary conditions, 
$\theta(0)=\pi,\, \theta(\infty)=0$.

Within a circular cell approximation 
the equilibrium parameters of Skyrmion lattices can be
derived by integration of Eq. (\ref{DifEq}) with
boundary conditions $\theta(0)=\pi,\, \theta(R)=0$
and a subsequent minimization of the lattice energy density
$W_{SL} =(2/R^2)\int^{R}_{0} w(\rho)\rho d\rho$
with respect to the cell radius $R$.\cite{JETP89,JMMM94}
Mathematically similar equations arise
for Skyrmion states in uniaxial noncentrosymmetric
ferromagnets.\cite{JMMM94}
In particular, there are crystal 
classes where the Dzyaloshinskii-Moriya
energy is described by Lifshitz invariants with gradients
only along the directions perpendicular to the unaxial
axis, e.g. for $D_{2d}$ classes,
$w_D = D (\Lambda_{yz}^{(x)}+\Lambda_{xz}^{(y)})$.
\cite{JETP89,PRB02}
This restriction of modulations to two dimensions proved to be 
crucial for the thermodynamical stability of the Skyrmion
states.\cite{JMMM94,condmat2010,PRB02}
In the high symmetry cubic helimagnets
the chiral energy 
$w_D = D\,\mathbf{M}\cdot \mathrm{rot}\mathbf{M}$ 
(\ref{density}) energetically favours 
the helical phases compared 
to the Skyrmion states.\cite{Nature06}

\begin{figure}
\includegraphics[width=8.5cm]{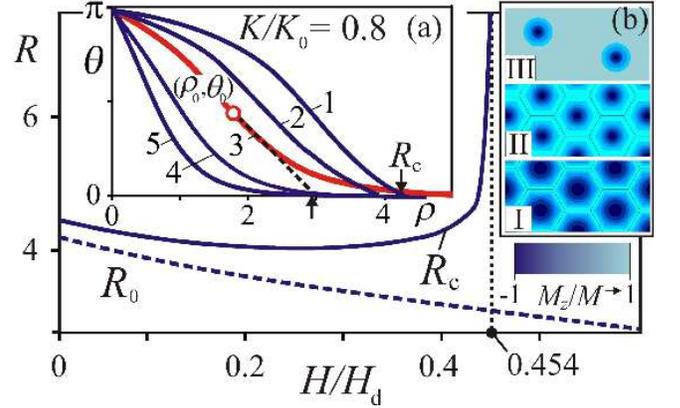}
\caption{
\label{Fig2}
(Color online) Equilibrium size $R_c$ of the Skyrmion
cell and the characteristic radius 
of the core $R_0$ as functions
of an applied magnetic field for $K/K_0 = 0.8$. 
Inset (a) shows the evolution of Skyrmion 
profiles $\theta (\rho)$ with increasing
magnetic field, the distributions of the perpendicular
magnetization ($M_z$) are sketched in Inset (b):
$H/H_d$ = 0 (1, I), 0.2 (2, II),
0.454 (3, III), 0.6 (4), 0.8 (5)
(profiles (1,2) describe the magnetization
in lattice cells (I, II), while profiles (4,5)
- isolated Skyrmions.
Profile (3) corresponds to the transition
of the Skyrmion lattice into a system
of isolated Skyrmions (III).
}
\end{figure} 
From the numerical investigation of Eq.~(\ref{DifEq}),
we now show that a sufficiently strong magnetic anisotropy $K$
stabilizes Skyrmionic textures 
in applied magnetic fields.
Typical solutions for $\theta(\rho)$ are 
plotted in the inset of Fig. \ref{Fig2}.
At zero field the Skyrmion cells have
a smooth distribution of the magnetization
(profile (1) in Fig. \ref{Fig2}, Inset).
An increasing  magnetic field gradually squeezes the
Skyrmion core (profiles (2),(3)) and transforms
the lattice into  a system of isolated Skyrmions
(profiles (4), (5)).
The Skyrmion core size can be introduced
in a manner commonly used 
for magnetic domain walls, \cite{Hubert74}
$R_0 = \rho_0 + \theta_0 /(d \theta /d \rho)_0 $
where $(\rho_0, \theta_0)$ is the inflection
point of profile $\theta(\rho)$ (Fig. \ref{Fig2}).
The plots in Fig.~\ref{Fig2} 
demonstrate a progressive localization of the
Skyrmion core $R_0$ accompanied by the expansion
of the lattice cell size $R_c$ with increasing $H$.

\begin{figure}
\includegraphics[width=8.5cm]{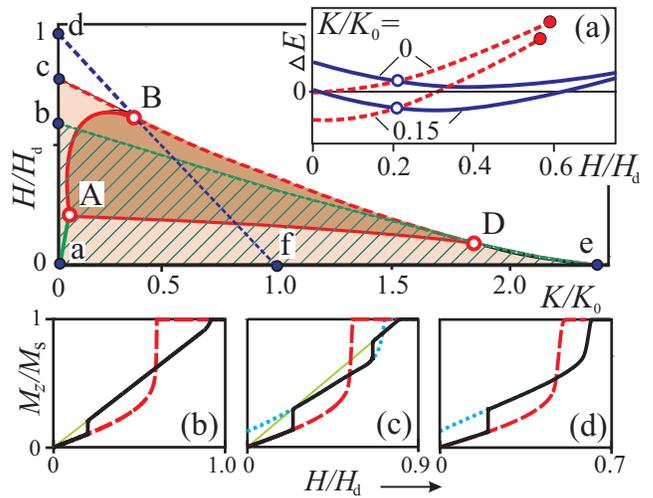}
\caption{
\label{PD}
(Color online). Magnetic phase diagram of the
solutions for model (\ref{density}).
Filled areas designate the regions of
global stability ($A-B-D$) and meta-stablity 
($a-b-c-e$) of Skyrmion lattices.
These regions transform by first-order processes
into the helocoids ($A-D$) or conical
phases ($A-B$).  
%
%The second-order transition
%into the saturatated state occurs
%at the critical line $B-D$.
%
The solutions for helicoids exist within
the area ($a-b-D-e$).
Inset (a) gives the differences $ \Delta E$
between the energies for the Skyrmion lattice 
and the conical phase (solid) and
the distorted and conical helices (dashed)
as functions of the applied field.
Insets (b), (c), (d) show magnetization 
curves  for different values of uniaxial
distortions: $K/K_0$ = 0.04 (b), 0.4 (c),
and  1.6 (d).
Solid lines indicate 
the globally stable phases, 
metastable states are shown 
by green thin (conical),
red dashed (helicoids),
blue dotted (Skyrmions) lines. 
}
\end{figure}

The functional (\ref{density}) includes
two independent control parameters, $K/K_0$ and $H/H_d$.
The calculated magnetic phase diagram 
in these variables (Fig.~\ref{PD}) 
provides a comprehensive analysis of model (\ref{density}).
%
%It was constructed by numerically solving
%the corresponding differential equations
%for the Skyrmion and helicoid states with
%a further comparison of the equilibrium energies
%for the conical ($W_{C}$ (\ref{conical})), helicoid
%$W_{H}$, and the Skyrmion lattice states $W_{SL}$. 
%
For $K$ = 0 
the conical phase is the globally 
stable state from zero field to the
saturation field ($0< H < H_d$),\cite{Bak80} 
while metastable solutions for the Skyrmion lattices and 
helicoids exist below the critical fields 
$ H_c/H_d = 0.8132 $ and
$H_b/H_d = \pi^2/16 = 0.6168$, correspondingly.\cite{JMMM94,Nature06}
%
%The solutions for the conical phase exist 
%below the critical line $(d-B-f)$.
%
The conical phase is the global minimum
of the system within the region ($a-A-B-d$)
and transforms discontinuously into
the Skyrmion (($A-B$)-line) and
helicoid phases (($a-A$)-line).
%
%The second-order flip into
%the saturated state occurs at the
%critical line ($d-B$) (Eq. (\ref{conical})).
%
A sufficiently strong $K$ suppresses 
the conical states, and only
modulations with the propagation
vectors perpendicular to the applied
field can exist (helicoids and Skyrmion lattices).
The Skyrmion states are thermodynamically stable
within a  curvilinear triangle $(A-B-D)$ with
vertices
$(A) = (0.050, 0.2158)$, 
$(B) = (0.3628, 0.6374)$,
and $(D) = (1.9004, 0.10)$) (Fig.~\ref{PD}).
The solutions for helicoids
exist within area ($a-b-D-e$).
%
%They are expressed 
%as a combination of elliptical functions \cite{Dz64}. 
%
%These describe the gradual ``unwinding'' of helices
%by an increasing applied field.
%
For a certain value of magnetic field 
the helicoid is transformed into 
the homogeneous state 
%by an unlimited growth of the
%period \cite{Dz64,PRB02}
at the critical line $(b-D-e)$
$D = (\pi/4) \sqrt{AK}[ \sqrt{1+\nu} 
+ \mathrm{arcsinh} (1/\sqrt{\nu}) ]$
where $\nu = H/(2KM)$.\cite{JMMM94}
Point ($e$) designates the
critical value of the anisotropy $K$ for 
the suppression of the modulated states in zero field 
($K_e/K_0 = \pi^2/4 = 2.467$).

The critical points $A$, $B$, $D$ separate
the phase diagram (Fig. \ref{PD}) into four
distinct regions. 
(I) In the low anisotropy
region ($K < K_A = 0.05 K_0$) only helical
states are realized as thermodynamically
stable phases.
%The magnetization curve 
%(Inset (b)) includes the first-order
%transition between the helicoid and conical
%phases and the second-order transition 
%from the conical to the saturated state.
%
(II) For $K_A < K < K_B = 0.363 K_0$ 
the Skyrmion lattice becomes absolutely
stable in a certain range of the applied field.
The Skyrmionic phase is separated from 
the helicoidal and conical states  
by first-order transitions.
(Inset(c)).
(III) For $ K_B < K < K_D = 1.90 K_0$
the magnetization curve includes a first-order
transition between the helicoid and the Skyrmion
lattice and the second-order transition of
the Skyrmion phase into the saturated state
(Inset (d)).
(IV) Finally for ( $K_D < K < K_e$)
the helicoids are thermodynamically stable
in the whole region where the modulated states
exist.
%
%The magnetization proccess includes the evolution
%of the helicoid into the saturated state.

% 
The equilibrium energies of the 
conical ($W_C$), helicoidal ($W_H$),
and Skyrmion ($W_{SL}$) phases
plotted as functions
$\Delta E_{SL(H)} = W_{SL(H)} - W_C$
versus the applied field
(Inset (a)  of Fig. \ref{PD}) 
help to elucidate the physical mechanisms
that lead to the formation of the different 
modulated states. 
The conical phase has a simple structure
with its single-harmonic rotation of the 
magnetization component 
$M_{\perp} = M \sin \theta$ (\ref{conical}). 
For zero $K$, i.e., for an isotropic helimagnet this 
provides a larger reduction of the energy density in an external field 
($w_D \propto -D M_{\perp}^2$) than
is possible for the (\textit{anharmonic}) modulations 
in the alternative phases of helicoids 
and Skyrmion lattices.
%
%The situation changes for finite uniaxial distortions $K$.
%
An increasing uniaxial anisotropy $K > 0$ 
gradually decreases $M_{\perp}$ (\ref{conical}) 
and the chiral energy contribution
in the energy of the conical phase. 
Correspondingly, the conical phase becomes
unstable with respect to the competing 
modulated states (Fig.~\ref{PD}).
%
%Generally a cubic helimagnet in the conical phase 
%behaves as a ferromagnet with the
%longitudinal susceptibility (\ref{conical})
%
%$\chi_{zz} = (M/H_D)(1- K/K_0)^{-1}$.
%
%With increasing $K$ the saturation field
%$H_c$ (\ref{conical}) bounding the 
%existence region of the conical phase
%decreases to zero at $K = K_0$.

The energetic advantage of Skyrmion
states is due to rotation
of the magnetization in two dimensions.
This (\textit{double-twist}) 
grants a larger reduction of the 
Dzyaloshinskii-Moriya energy than 
a single-direction rotation in helical phases.
Thus, the double-twist yields a lower energy density 
in the Skyrmion cores compared to 
helical states.
On the other hand, the incompatibility 
of spin configurations
near the edges of the hexagonal cells 
leads to an excess of the energy
density in this region.\cite{Nature06}
The analysis shows that at zero field this 
energy cost outweighs the energy 
gain in the Skyrmion core.
An increasing external magnetic 
field anti-parallel to the magnetization
in the Skyrmion center gradually 
decreases the total energy by
suppressing the energy cost near the wall-like structure 
surrounding the Skyrmion cores with the shape of 
a honeycomb (Fig. \ref{Fig2}, Inset). 
At a finite field as marked by
the $A-D$ line in Fig. \ref{PD},
the Skyrmion lattice has lower
energy than the alternative
helical states. Because the topology
of the helix and the Skyrmion lattices are different,
this field-driven transition has to take place 
by a first-order process.
Thus, \textit{in uniaxially distorted 
cubic helimagnets the thermodynamical stability
of Skyrmion lattices is reached
as a combined effect of applied
magnetic field, that causes the localization
of the cell core, and the uniaxial anisotropy
to suppress the alternative conical states}.

Skyrmionic states now have been observed
in nanolayers of Fe$_{0.5}$Co$_{0.5}$Si \cite{Yu10}.
To explain the stability of the Skyrmion phase
in this system, we argue that
surface-induced uniaxial anisotropy  
in this system suppresses the cone 
phase and stabilizes skyrmion and 
helicoid modulations in crystal plates 
that are thin enough.
The magnetic transformation under field 
reported for the Fe$_{0.5}$Co$_{0.5}$Si 
films show a first-order process 
from helicoids at low field 
into the dense Skyrmion phase.
At high fields isolated  Skyrmions are set free 
and form disordered ensembles.
This sequence of magnetization processes corresponds to
the calculated behavior for systems
with intermediate uniaxial anisotropy, Fig.~3(d).
As seen in Ref.~\cite{Yu10}, 
there is a type of melting of the ordered 
Skyrmion lattice at higher fields 
and temperatures where the free 
Skyrmions form weakly coupled 
disordered aggregates. 
This observations corresponds to the existence region 
of free Skyrmions in the magnetic phase diagram above 
the stability range for dense stable Skyrmion lattices,
which is a feature of the generic phase
diagrams for Skyrmion phases calculated in Ref.~\cite{condmat2010}.
Thus, the experimental observations 
reported in Ref.~\cite{Yu10},
are in close agreement with theoretical predictions
on the behavior of Skyrmionic phases, 
as composed of particle-like radial objects 
(in two spatial dimensions),
that remain intrinsically stable beyond the 
existence range of lattice-like condensed 
phases.\cite{JMMM94,Nature06,condmat2010} 

For MnSi, earlier experiments
\cite{FranusMuir84}
and analysis \cite{PlumerWalker82} of magnetoelastic couplings
allow a quantitative estimate showing that the effects 
predicted here can be achieved in experiment.
The magnetoelastic coupling with uniaxial strains $u_{zz}$
is given by \(w_{me}=b \,u_{zz}\,(M_z/M_S)^2\),
where $M_S=$~50.9~A/m is the saturation magnetization \cite{Bloch75} 
and $b=$~7.4~GPa  is a magnetoelastic coefficient 
derived from the magnetostriction data in Ref.~\onlinecite{FranusMuir84}.
Using exchange constant $A=$~0.11~pJ/m,
as estimated from the spin-wave stiffness
reported in Ref. \cite{Grigoriev05},
and $D=2\,q_0\,A=$~0.86~$\mu$J/m$^2$ for MnSi 
\cite{Nature06}
we have $K_0\simeq$~17~kJ/m$^3$ and a dimensionless 
scale $b/K_0\simeq$~44 for the induced anisotropy.
Thus, a modest strain $u_{zz}=$~0.0024
is sufficient to reach an induced anisotropy $K/K_0=$~0.1 well 
within the region for stable 
Skyrmion lattices in the phase diagram Fig.~\ref{PD}.
This strain corresponds to a tensile stress $\sigma_{zz}$=~680 MPa for MnSi 
by using the elastic constant $c_{11}$=~283~GPa.\cite{Stishov08}
The rather low uniaxial stress necessary to stabilize the Skyrmion lattice 
is particularly relevant for pressure experiments
with a uniaxial disbalance of the applied stresses, but it could also be 
achieved in epitaxial films.
Finally, intrinsic anisotropy $w_a$ (\ref{density})
also makes a contribution to stabilize Skyrmionic states
as both exchange ($B$) and cubic ($K_c$)
anisotropies violate the ideal spin configuration of
the cone  (\ref{conical}) and \textit{increases} the energy
of this phase \cite{condmat2010}.
This favours Skyrmion lattices for certain directions of 
the applied field. Particularly, in MnSi ($B < 0$) 
for Skyrmion lattices oriented along [001] type axes 
$\Delta E_{SL} (H/H_d)$ = 0 at critical point
$(B_{cr}= 0.10 A/D^2, H_{cr} = 0.408 H_d)$.
Thus, for  $|B| > B_{cr}$ Skyrmion lattices are globally 
stable in a certain interval 
of magnetic fields around $H_{cr}$.

In conclusion, we have shown that in cubic helimagnets 
uniaxial distortions effectively suppress helical states
and stabilize the Skyrmion states in a broad range of the
applied fields.

\begin{acknowledgments}
The authors thank S. Bl\"ugel, S. Stishov, and H. Wilhelm
for discussions.
Support by DFG project RO 2238/9-1 is gratefully acknowleged.
\end{acknowledgments}


\begin{thebibliography} {99}
\bibitem{Remark1}
%
Energy terms of the fourth order in gradients
of the magnetization arise as a continuum limit
of discretized models with competing exchange interactions 
\cite{Hubert74}.
%
These systems are characterized by
short modulation periods with spontaneously 
broken chiral  symmetry.
\end{thebibliography}
\end{document}